\documentclass[12pt,showpacs,showkeys]{revtex4}
\usepackage{amssymb}
\usepackage{amsmath}
\usepackage{graphicx}
\usepackage{amsbsy}

\begin{document}

\title{Multicomponent adhesive hard sphere models\\
and short-ranged attractive interactions\\
in colloidal or micellar solutions }
\author{Domenico Gazzillo$^{\S }$, Achille Giacometti$^{\S }$, Riccardo
Fantoni$^{\S }$, and Peter Sollich$^{\ddagger }$}
\email{gazzillo@unive.it}
\affiliation{$^{^{\S }}$Istituto Nazionale per la Fisica della Materia and Dipartimento
di Chimica Fisica, Universit\`{a} di Venezia, S. Marta DD 2137, I-30123
Venezia, Italy\\
$^{\ddagger }$King's College London, Department of Mathematics, Strand,
London WC2R 2LS, UK}
\date{\today }

\begin{abstract}
We investigate the dependence of the stickiness parameters $t_{ij}=1/(12\tau
_{ij})$ -- where the $\tau _{ij}$ are the conventional Baxter parameters --
on the solute diameters $\sigma _{i}$ and $\sigma _{j}$ in multicomponent
sticky hard sphere (SHS) models for fluid mixtures of mesoscopic neutral
particles. A variety of simple but realistic interaction potentials,
utilized in the literature to model \textit{short-ranged attractions}
present in real solutions of colloids or reverse micelles, is reviewed. We
consider: i) van der Waals attractions, ii) hard-sphere-depletion forces,
iii) polymer-coated colloids, iv) solvation effects (in particular
hydrophobic bonding and attractions between reverse micelles of water-in-oil
microemulsions). We map each of these potentials onto an equivalent SHS
model, by requiring the equality of the second virial coefficients. The main
finding is that, for most of the potentials considered, the size-dependence
of $t_{ij}(T,\sigma _{i},\sigma _{j})$ can be approximated by essentially
the same expression, i.e.\ a simple polynomial in the variable $\sigma
_{i}\sigma _{j}/\sigma _{ij}^{2}$, with coefficients depending on the
temperature $T$, or -- for depletion interactions -- on the packing fraction 
$\eta _{0}$ of the depletant particles.
\end{abstract}

\maketitle

\newpage

\section{Introduction}

Theoretical investigation of solutions of \textit{mesoscopic} particles -
with sizes within the range 10 - 10$^{4}$ \AA\ - such as colloids, micelles
and globular proteins, is more problematic than the study of fluids with
atomic or simple molecular constituents - with sizes within the range 1-10 
\AA\ \cite{Barrat03,Lyklema00,Frenkel00,Kleman03}. The main difficulties are
due to the large difference between solute and solvent molecular sizes, as
well to the possible presence of high electric charges and large
charge-asymmetries. Treating mixtures of macroions and microions, with
strong long-ranged Coulombic forces, represents a challenge for the most
typical methods of the modern statistical-mechanical theory of fluids,
namely Monte Carlo (MC) or molecular dynamics (MD) computer simulations and
integral equations (IE) based on the Ornstein-Zernike equation coupled with
approximate \textquotedblleft closures \textquotedblright\ \cite{Hansen86}.
Large size-asymmetries entail very different time scales in MD simulations
and may lead to ergodicity problems both in MC and MD calculations.
Moreover, large size-differences imply several difficulties even when using
IE theories.

For simplicity, the present paper will be restricted to fluids of \textit{%
neutral} particles with spherically symmetric interactions, neglecting all
Coulombic forces due to net electric charges. Starting from a fluid mixture
with one or more solute species (big particles, or macroparticles) and one
\textquotedblleft solvent\textquotedblright\ species (much smaller molecules
or microparticles, which might be either a true solvent or polymer coils,
smaller colloidal particles, etc.), we will adopt an \textit{effective fluid}
approach, which eliminates all large size-asymmetries by averaging out the
microscopic degrees of freedom corresponding to the solvent \cite%
{Amokrane04,Louis02}. As a consequence, the influence of the solvent is
incorporated into an\textit{\ effective} potential for the interaction
between big particles, and the initial mixture is reduced to a fluid made up
of only solute molecules (one or more components). Usually, at the simplest
level of description the effective potential includes, in addition to a
steeply repulsive part, a very \textit{short-ranged attractive} one, whose
range is a small fraction of the macroparticle size. Recall that a force is
said to be \textquotedblleft short-ranged\textquotedblright\ if it derives
from a potential $\phi _{ij}\left( r\right) $ which vanishes as $r^{-n}$
with $n\geq 4$ when $r\rightarrow \infty $ \cite{Widom02,Friedman85}; the
force $-\partial \phi _{ij}/\partial r$ then decays as $r^{-\left(
n+1\right) }$. This definition of short-ranged potentials is clearly related
to the second virial coefficient $B_{2,ij},$ which is a central quantity in
our paper: when the forces are short-ranged in the above-mentioned sense,
the integral which defines $B_{2,ij}$ (see Eq. (\ref{v2}) below) is finite,
whereas it diverges for long-ranged interactions, i.e.\ when $r\leq
3$. Note that
the definition of short-ranged forces is not unique in the literature. For instance,
in Hirschfelder's classical reference book \ \cite{Hirschfelder65}
short-range forces are the \textquotedblleft valence or chemical
forces\textquotedblright , arising from overlap of electron clouds at very
short intermolecular separations. The potential of such repulsive, and often
highly directional, forces varies exponentially with the distance $r$. On
the other hand, all potentials proportional to inverse powers of $r$ are
called \textquotedblleft long-ranged\textquotedblright\ by Hirschfelder \cite%
{Note1}.

Once a reasonable approximation to the effective potential is known, it
could be employed in both computer simulations or IE calculations.
Unfortunately, IEs can be solved analytically only in very specific cases,
for some potentials and within particular \textquotedblleft
closures\textquotedblright\ \cite{Hansen86}. The simplest model with both
repulsion and attraction which is analytically tractable refers to a fluid
made up of hard spheres (HS) with an infinitely narrow and infinitely deep
attractive tail. This highly idealized model of \textit{adhesive} or \textit{%
sticky hard spheres }(SHS) was proposed by Baxter \cite{Baxter6871}, and
admits an analytical solution within the Percus-Yevick (PY) approximation 
\cite{Baxter6871,Perram75,Barboy79}. Notwithstanding its crudeness and known
shortcomings \cite{Stell91}, the SHS model is not a purely academic
exercise. In fact, it has seen continuously growing interest in the last two
decades, because of its ability to describe semiquantitatively many
properties of real fluids of neutral spherical particles, such as colloidal
suspensions, micelles, protein solutions, microemulsions, and systems
exhibiting phase transitions of several types (see for example Refs. \cite%
{Stell91,Regnaut89,Jamnik9601,Santos98} and references therein). Accurate
simulation data for one-component SHS have recently been reported by Miller
and Frenkel \cite{Miller0304}.

Because of the simplicity of the SHS model, it has often been suggested to
model potentials comprising a hard core and short-ranged attractive tail by
means of sticky potentials. To achieve this one needs to define an
appropriate \textit{equivalence} between the actual interaction and its 
\textit{sticky representation}. This \textit{mapping} of a generic
short-ranged potential onto a SHS interaction is usually accomplished by
requiring the two different models to have equal second virial coefficients 
\cite{Regnaut89,Regnaut95}. Moreover, when applied to mixtures, this
approach requires a further step, and this is the main point addressed in
the present work.

In a series of earlier papers \cite%
{Gazzillo00,Gazzillo02,Gazzillo03,Gazzillo04,Gazzillo05}, we investigated
the multi-component SHS model, focusing on its possible application to 
\textit{polydisperse} colloidal suspensions, namely to mixtures where the
number $p$ of components is so \textit{large }that it can effectively be
regarded as stemming from a continuous distribution. This is, for instance,
the case of size polydispersity, where -- in the discrete notation -- a SHS
mixture is fully characterized by two sets of parameters, i.e., the HS
diameters $\left\{ \sigma _{i}\right\} $ and the \textquotedblleft
stickiness\textquotedblright\ coefficients $\left\{ t_{ij}=1/(12\tau
_{ij})\right\} $ ($\tau _{ij}$ are Baxter's parameters); the latter depend
on temperature $T$ and the strength of the interparticle adhesion.
Intuitively, one expects $t_{ij}$ to depend on the diameters $\sigma _{i}$
and $\sigma _{j}$ of the interacting particles $i$ and $j$, but it is not
easy to specify \textit{a priori} the correct functional form, and in our
previous papers we attempted some reasonably motivated choices for such a
dependence.

The main purpose of the present paper is to investigate the relationship
between stickiness coefficients and particles sizes, and thus to get new
insights into the possible forms of the function $t_{ij}=t_{ij}\left(
T,\sigma _{i},\sigma _{j}\right) ,$ starting from a physically sound basis.
To achieve this, we will present an overview of the most important
short-ranged attractive interactions occurring in real solutions of colloids
or micelles. In doing this, our claim is not to be fully exhaustive, but
rather to gather sufficient physical information about the mechanisms which
cause short-ranged attractive interactions in solutions of mesoscopic
particles, and the corresponding simplest model potentials used for their
representation.

By considering several different systems -- dispersion forces, depletion
forces, polymer-coated colloids, solvation forces (in particular,
hydrophobic interactions, and reverse micelles in water-in-oil
microemulsions) -- we have surprisingly found strong similarities among the
simplest models employed to represent this wide variety of physical
phenomena. By constructing, for each of the relevant potentials, an
equivalent SHS representation, we will deduce and compare the corresponding
expressions for $t_{ij}=t_{ij}\left( T,\sigma _{i},\sigma _{j}\right) $.

The paper is organized as follows. In Section II we will introduce the basic
formalism, concerning the second virial coefficient, the Baxter SHS model,
and the mapping rule for getting the equivalent SHS potential from a given
short-ranged attraction. Sections III is dedicated to the direct van der
Waals interaction, while Sections IV, V and VI survey the most important
short-ranged attractions that are indirect, i.e.\ mediated by the solvent.
The hydrophobic effect and interactions between reverse micelles will be
considered in Section VI, as particular cases of solvation forces. For each
model potential, a reasonable approximation to the corresponding $%
t_{ij}=t_{ij}\left( T,\sigma _{i},\sigma _{j}\right) $ will be calculated.
Finally, a summary, with a brief discussion, and our conclusions will be
given in Section VII.

\section{Basic formalism}

\subsection{Second virial coefficient}

For a multi-component fluid, the second virial coefficient of the osmotic
pressure reads %$B_{2}=\dsum\limits_{i,j}x_{i}x_{j}B_{2,ij}$ , 
$B_{2}=\sum_{i,j}x_{i}x_{j}B_{2,ij}$ , where $x_{i}$ is the molar fraction
of species $i$, and the partial second virial coefficient for the $i-j$
interaction is given by%
\begin{equation}
B_{2,ij}=-\frac{1}{2}\int f_{ij}(r)\ d\mathbf{r}=-2\pi \int_{0}^{+\infty
}f_{ij}(r)r^{2}dr,  \label{v2}
\end{equation}%
with%
\begin{equation}
f_{ij}(r)=\exp \left[ -\beta \phi _{ij}\left( r\right) \right] -1  \label{v3}
\end{equation}%
being the Mayer function, $\beta =\left( k_{B}T\right) ^{-1}$, $k_{B}$ the
Boltzmann constant, and $T$ the absolute temperature.

When the actual potential consists of a hard core plus a short-ranged
attractive tail, i.e., $\phi _{ij}\left( r\right) =\phi _{ij}^{\mathrm{HS}%
}\left( r\right) +\phi _{ij}^{\mathrm{tail}}\left( r\right) $, one gets 
\begin{equation}
B_{2,ij}=B_{2,ij}^{\mathrm{HS}}+B_{2,ij}^{\mathrm{tail}}  \label{v4}
\end{equation}
\begin{equation}
B_{2,ij}^{\mathrm{tail}}=-2\pi \int_{\sigma _{ij}}^{+\infty }f_{ij}^{\mathrm{%
tail}}(r)r^{2}dr=B_{2,ij}^{\mathrm{HS}}\left[ -3\int_{1}^{+\infty }f_{ij}^{%
\mathrm{tail}}(\sigma _{ij}x)x^{2}dx\right] ,  \label{v5}
\end{equation}%
where $\sigma _{i}$ is the HS diameter for particles of species $i$ and we
set $\sigma _{ij}~=\left( \sigma _{i}+\sigma _{j}\right) /2$ as usual,
introducing also the shorthands $B_{2,ij}^{\mathrm{HS}}=\left( 2\pi
/3\right) \sigma _{ij}^{3}$ and $f_{ij}^{\mathrm{tail}}(r)=\exp \left[
-\beta \phi _{ij}^{\mathrm{tail}}\left( r\right) \right] -1$.

Often, the required integration cannot be performed analytically, but if $%
\phi _{ij}^{\mathrm{tail}}(r)$ is sufficiently small compared to the thermal
energy $k_{B}T$, then approximate analytical expressions may be obtained
after expanding the Mayer function $f_{ij}^{\mathrm{tail}}(r)$ in powers of $%
\ Y\equiv -\beta \phi _{ij}^{\mathrm{tail}}\left( r\right) $. A numerical
estimate of the range of applicability and the maximum relative error $%
\Delta _{\mathrm{\max }}=\max \left\vert 1-f_{\mathrm{approx}}/f\right\vert $%
, for each of the three simplest approximations, is 
\begin{equation}
f=e^{Y}-1\approx \left\{ 
\begin{array}{ccccc}
Y &  & 0<Y\lesssim 0.1 &  & \Delta _{\mathrm{\max }}\sim 5\% \\ 
Y+Y^{2}/2 &  & 0<Y\lesssim 0.6 &  & \Delta _{\mathrm{\max }}\sim 5\% \\ 
Y+Y^{2}/2+Y^{3}/6 &  & 0<Y\lesssim 1 &  & \Delta _{\mathrm{\max }}\lesssim
3\%\ .%
\end{array}%
\right.  \label{v5b}
\end{equation}

\subsection{Adhesive hard spheres as a limiting case of square-well model}

Probably, the simplest two-parameter representation of a spherically
symmetric interaction with steeply repulsive core and short-ranged
attractive tail is the \textit{square-well} (SW) potential%
\begin{equation}
\phi _{ij}^{\mathrm{SW}}(r)=\left\{ 
\begin{array}{ll}
+\infty & \qquad 0<r<\sigma _{ij}~, \\ 
-\epsilon _{ij} & \qquad \sigma _{ij}\leq r\leq \sigma _{ij}+w_{ij}~, \\ 
0 & \qquad r>\sigma _{ij}+w_{ij}~,%
\end{array}%
\right.  \label{b1}
\end{equation}%
with $\epsilon _{ij}>0$ and $w_{ij}$ being the depth and width of the well,
respectively. The corresponding partial second virial coefficient reads 
\begin{eqnarray}
B_{2,ij}^{\mathrm{SW}} &=&B_{2,ij}^{\mathrm{HS}}\left\{ 1-\left( e^{\beta
\epsilon _{ij}}-1\right) \left[ \left( 1+\Delta _{ij}\right) ^{3}-1\right]
\right\}  \label{b2} \\
&=&B_{2,ij}^{\mathrm{HS}}\left[ 1-3\left( e^{\beta \epsilon _{ij}}-1\right)
\left( \Delta _{ij}+\Delta _{ij}^{2}+\frac{1}{3}\Delta _{ij}^{3}\right) %
\right]  \notag
\end{eqnarray}%
with $\Delta _{ij}=w_{ij}/\sigma _{ij}\geq 0$. Eq. (\ref{b2}) shows that, if
the well is narrow $\left( \Delta _{ij}\ll 1\right) $, $B_{2,ij}^{\mathrm{SW}%
}$ can be significantly different from $B_{2,ij}^{\mathrm{HS}}$ only when
the attraction is strong enough $\left( e^{\beta \epsilon _{ij}}\gg 1\right)
.$

Unfortunately, despite the simplicity of the SW model, no satisfactory
analytical solution of the resulting IEs has been found so far. However,
such a solution \emph{can} be found within the Percus-Yevick (PY)
approximation for a special limiting case, when the well width $\Delta _{ij}$
goes to zero but the depth $\epsilon _{ij}$ goes to infinity in such a way
that the contribution of the attraction to the second virial coefficient remains
finite and different from zero (Baxter's sticky limit) \cite{Baxter6871}.
The short-ranged attraction becomes a surface adhesion, and the particles of
the resulting model are thus named \textit{adhesive} or \textit{sticky hard
spheres}. From Eq. (\ref{b2}) one sees that Baxter's condition on $B_{2,ij}^{%
\mathrm{SW}}$ requires the product $\left( e^{\beta \epsilon _{ij}}-1\right)
\Delta _{ij}\equiv t_{ij\text{ \ }}$to be independent of $\Delta _{ij}$ for
small $\Delta _{ij}$, and this leads to the following condition for the SW
depth 
\begin{equation}
\epsilon _{ij}^{\ \mathrm{Baxter}\text{ }\mathrm{SW}}=k_{B}T\ \ln \left( 1+%
\frac{t_{ij}}{\Delta _{ij}}\right) .  \label{b3}
\end{equation}%
As previously mentioned, our $t_{ij}$ is simply related to Baxter's original
parameter $\tau _{ij}$ by 
\begin{equation}
t_{ij}=\frac{1}{12\tau _{ij}}\geq 0\ .  \label{b4}
\end{equation}%
Here, $t_{ij}$ measures the strength of surface adhesiveness or
\textquotedblleft stickiness\textquotedblright\ between particles of species 
$i$ and $j$, and must be an unspecified \textit{decreasing} function of $T.$
In fact, as $T\rightarrow \infty $ one must also have $\tau _{ij}\rightarrow
\infty $, in order to recover the correct HS limit. The SHS models must
therefore satisfy the \textit{high-temperature condition} 
\begin{equation}
\lim_{T\rightarrow \infty }t_{ij}=0.  \label{b5}
\end{equation}

A consequence of Eq. (\ref{b3}) is a very simple expression for the SW Mayer
function 
\begin{equation}
f_{ij}^{\ \mathrm{Baxter}\text{ }\mathrm{SW}}(r)=\left\{ 
\begin{array}{ll}
-1 & \qquad 0<r<\sigma _{ij}~, \\ 
t_{ij}\ \sigma _{ij}/w_{ij} & \qquad \sigma _{ij}\leq r\leq \sigma
_{ij}+w_{ij}~, \\ 
0 & \qquad r>\sigma _{ij}+w_{ij}~,%
\end{array}%
\right.  \label{b6}
\end{equation}%
Baxter focused on $f_{ij}$, since this quantity directly determines $%
B_{2,ij} $ and, furthermore, the coefficients in the cluster expansion of
thermodynamic properties and correlation functions can be expressed in terms
of multi-dimensional integrals of products of Mayer functions \cite{Hansen86}%
. The simple functional form of $f_{ij}^{\ \mathrm{Baxter}\text{ }\mathrm{SW}%
}(r)$ then allows one to calculate analytically many quantities of interest.
In the \textquotedblleft sticky limit\textquotedblright\ $\left\{
w_{ij}\right\} \rightarrow \left\{ 0\right\} $, the Mayer function becomes%
\begin{equation}
f_{ij}^{\ \mathrm{SHS}}(r)=\left[ \theta \left( r-\sigma _{ij}\right) -1%
\right] +t_{ij}\ \sigma _{ij}\ \delta _{+}\left( r-\sigma _{ij}\right)
\label{b7}
\end{equation}%
with $\theta (x)$ being the Heaviside function ($=0$ when $x<0$, and $=1$
when $x>0$) and $\delta _{+}(x)$ an asymmetric Dirac distribution \cite%
{note1}, while the SHS second virial coefficient is simply 
\begin{equation}
B_{2,ij}^{\mathrm{SHS}}=B_{2,ij}^{\mathrm{HS}}\left( 1-3t_{ij}\right) .
\label{b8}
\end{equation}

\subsection{Mapping onto equivalent SHS model}

On comparing Eqs. (\ref{b8}) and (\ref{v5}), one has%
\begin{equation}
t_{ij}^{\mathrm{eq}\left( \mathrm{tail}\right) }=-\frac{B_{2,ij}^{\mathrm{%
tail}}}{3B_{2,ij}^{\mathrm{HS}}}\ ,
\end{equation}%
and hence the following \textit{mapping rule}: the parameters $t_{ij}$ of
the \textit{equivalent} SHS model must be given by 
\begin{equation}
t_{ij}^{\mathrm{eq}\left( \mathrm{tail}\right) }=\frac{1}{\sigma _{ij}^{3}}%
\int_{\sigma _{ij}}^{+\infty }f_{ij}^{\mathrm{tail}}(r)r^{2}dr=\int_{1}^{+%
\infty }f_{ij}^{\mathrm{tail}}(\sigma _{ij}x)x^{2}dx.  \label{b10}
\end{equation}%
This is the main relation used in the remaining part of the paper. The
superscript in $t_{ij}^{\mathrm{eq}\left( \mathrm{tail}\right) }$ means:
this $t_{ij}$ yields \textit{the SHS potential equivalent to }$\phi _{ij}^{%
\mathrm{tail}}$.

\section{van der Waals attraction}

The main \textit{direct }attraction between two \textit{neutral }molecules $%
i $ and $j$ is the \textit{van der Waals} (vdW) interaction, represented by
the potential $\phi _{ij}^{\mathrm{vdW}}(r)=-C_{ij}^{\mathrm{vdW}}r^{-6}$,
which is - in general - the sum of three different contributions. For most
simple molecules \ - except the small highly polar ones - the vdW attraction
is almost exclusively determined by the dispersion forces; the latter are in
fact the only contribution to the vdW forces if both molecules are nonpolar.

\subsection{Dispersion forces}

The \textit{dispersion }or \textit{London} forces are \textit{%
induced-dipole/induced-dipole} interactions, whose potential is given by the
London formula \cite{Hirschfelder65} 
\begin{equation}
\phi _{ij}^{\mathrm{disp}}(r)=-\frac{C_{ij}}{r^{6}}\ ,\qquad C_{ij}=\frac{3}{%
2}\frac{I_{i}I_{j}}{I_{i}+I_{j}}\ \alpha _{i}^{\prime }\alpha _{j}^{\prime
}\ ,\qquad \text{for large }r,  \label{b13}
\end{equation}%
where $I_{i}$ and $\alpha _{i}^{\prime }$ are, respectively, the ionization
energy and \textit{polarizability volume} for molecules of species $i$. As
the name suggests, $\alpha _{i}^{\prime }$ has the dimensions of volume. It
can also be written as $\alpha _{i}/\left( 4\pi \varepsilon _{0}\right) $,
where $\varepsilon _{0}$ is the permittivity of the vacuum and $\alpha _{i}$
is the polarizability of species $i$, which increases with increasing
molecular size and number of electrons. Hence the polarizability volume is
proportional to the molecular volume, i.e., $\alpha _{i}^{\prime }\propto
\sigma _{i}^{3}$.

This \textit{polarizability effect} alone can produce considerable molecular
attraction, and is responsible for the formation of liquid phases from gases
of nonpolar substances (argon, hydrogen, nitrogen, etc.). The name
\textquotedblleft dispersion forces\textquotedblright\ stems from the fact
that the electronic oscillations producing the London attraction are also
responsible for the dispersion of light.

\subsection{Hamaker's macroscopic approximation}

Colloids, micelles and globular proteins are mesoscopic particles formed by
a very large number of polarizable molecules (typically $10^{10\text{ }}$in
micrometer-sized particles) \cite{Barrat03}. As a consequence, the total
attraction energy between such macroparticles can be obtained by pairwise
summation of London energies between all molecules of the two interacting
bodies. Hamaker \cite{Hamaker37} performed an approximate calculation \cite%
{Lyklema00} for the energy of interaction of two fully \textit{macroscopic}
bodies $i$ and $j$ in a vacuum, with densities $\rho _{i}$ and $\rho _{j}$
and occupying volumes $V_{i}$ and $V_{j}$. Replacing the discrete
distribution of molecules inside each body with a continuous one, Hamaker
obtained for two spheres of arbitrary size \cite{Lyklema00} 
\begin{eqnarray}
\phi _{ij}^{\mathrm{H}}\left( r\right) &=&-\frac{A_{ij}^{\mathrm{H}}}{12}%
\left[ \frac{\sigma _{i}\sigma _{j}}{r^{2}-\sigma _{ij}^{2}}+\frac{\sigma
_{i}\sigma _{j}}{r^{2}-L_{ij}^{2}}+2\ln \left( \frac{r^{2}-\sigma _{ij}^{2}}{%
r^{2}-L_{ij}^{2}}\right) \right]  \notag  \label{b18} \\
&&  \notag \\
&=&-\frac{A_{ij}^{\mathrm{H}}}{12}\left[ \frac{\sigma _{i}\sigma _{j}}{r^{2}}%
\left( \frac{1}{1-\sigma _{ij}^{2}/r^{2}}+\frac{1}{1-L_{ij}^{2}/r^{2}}%
\right) +2\ln \left( \frac{1-\sigma _{ij}^{2}/r^{2}}{1-L_{ij}^{2}/r^{2}}%
\right) \right]  \label{b18b}
\end{eqnarray}%
where $L_{ij}=\left\vert \sigma _{i}-\sigma _{j}\right\vert /2,$ and $\sigma
_{ij}<r<+\infty .$ Here, $A_{ij}^{\mathrm{H}}=\pi ^{2}\rho _{i}\rho
_{j}C_{ij}$ \cite{Kleman03} is referred to as Hamaker's constant, and has
dimensions of energy. As $C_{ij}\propto \alpha _{i}^{\prime }\alpha
_{j}^{\prime }\propto \sigma _{i}^{3}\sigma _{j}^{3}$, and $\rho _{i}\rho
_{j}\propto \sigma _{i}^{-3}\sigma _{j}^{-3}$, $A_{ij}^{\mathrm{H}}$ is
nearly independent of $i$ and $j$. In the case where all mesoscopic
particles are made up of the same material but have different diameters
(discrete size polydispersity) $A_{ij}^{\mathrm{H}}$ reduces to $A_{\mathrm{H%
}}=\pi ^{2}\rho ^{2}C$, which is a property of the material itself.

Hamaker's macroscopic result has also been applied to mesoscopic particles,
with the justification that the potential (\ref{b18b}) has a \textit{scaling
property}: if $r,\sigma _{i},\sigma _{j}$ are all multiplied by a factor $%
\gamma ,$ the attraction energy remains unaltered, i.e., $\phi _{ij}^{%
\mathrm{H}}\left( \gamma r,\gamma \sigma _{i},\gamma \sigma _{j}\right)
=\phi _{ij}^{\mathrm{H}}\left( r,\sigma _{i},\sigma _{j}\right) $. Note,
however, that Hamaker's formula refers to two spheres in free space, i.e.,
it neglects the screening of London forces due to the suspending medium.

In the limit $r\rightarrow +\infty $, a series expansion of Eq. (\ref{b18b})
yields 
\begin{equation}
\phi _{ij}^{\mathrm{H}}\left( r\right) \approx -\frac{A_{ij}^{\mathrm{H}}}{36%
}\frac{\ \sigma _{i}^{3}\sigma _{j}^{3}}{r^{6}},\text{\qquad for }r\gg
\sigma _{ij\ }>L_{ij}\text{\ ,}  \label{b18c}
\end{equation}%
which means that at large distances the two spheres behave, to leading
order, like point-particles even though the factors $\sigma _{i}^{3}$ and $%
\sigma _{j}^{3}$ stem from HS volumes.

On the other hand, the Hamaker potential is singular at contact, i.e., when $%
r\rightarrow \sigma _{ij}$. This is due to the approximation of regarding
the two spheres as continuous distributions of \textit{point-particles},
neglecting all intermolecular repulsions. The leading divergence is 
\begin{equation}
\phi _{ij}^{\mathrm{H}}\left( r\right) \approx -\frac{A_{ij}^{\mathrm{H}}}{12%
}\frac{\sigma _{i}\sigma _{j}}{r^{2}-\sigma _{ij}^{2}}\approx -\frac{A_{ij}^{%
\mathrm{H}}}{24}\frac{\sigma _{i}\sigma _{j}}{\sigma _{ij}}\frac{1}{r-\sigma
_{ij}}\text{ \ \ \ \ \ \ for \ \ }0<r-\sigma _{ij}\ll \min \left( \sigma
_{i}{},\sigma _{j}\right) .  \label{b19}
\end{equation}%
This divergence simply means that the continuum picture must break down and
molecular granularity - with excluded-volume effects - cannot be neglected
once the closest distance $r-\sigma _{ij}$ between the two spherical
surfaces becomes very small.

Such a deep attractive potential would lead to irreversible association or
\textquotedblleft flocculation\textquotedblright\ of the suspended
particle. This effect can be avoided in one of two different ways, namely
by \textit{charge stabilization} or \textit{steric stabilization}. In the
first case, some surface chemical groups of the particles become partially
ionized in water, and the resulting electrostatic repulsion makes close
contact impossible. In the second case, stabilization is achieved by
grafting polymer chains (\textquotedblleft hair\textquotedblright ) to the
particle surfaces. Both stabilization mechanisms - extensively used for
colloidal suspensions - imply that the closest approach distance between $i$
and $j$ becomes larger than $\sigma _{ij}$, i.e., $\sigma _{ij}^{\mathrm{eff}%
}=\sigma _{ij}+\delta $, with $\delta >0$ being an additional characteristic
length. The Hamaker singularity at contact is thus avoided, and the vdW
attraction may then be treated as a small \textit{perturbation}, if the
effective HS diameter is sufficiently large compared to the bare one (in
sterically stabilized colloidal suspensions, $\sigma ^{\mathrm{eff}}$
exceeds $\sigma $ typically by $10\%$). Moreover, it is possible to strongly
reduce the value of the Hamaker constant by \textquotedblleft refractive
index matching\textquotedblright\ \cite{Lyklema00}.

A numerical estimate of the strength of the Hamaker attraction is given - in the
one-component case, for simplicity - by the quantity%
\begin{equation}
Y_{\mathrm{\max }}\equiv -\beta \phi ^{\mathrm{H}}\left( \sigma +\delta
\right) =3\ H\left( 1+\lambda \right) \frac{T_{\mathrm{H}}}{T},  \label{b20}
\end{equation}%
where $\lambda \equiv \delta /\sigma $, 
\begin{equation}
H(u)=\frac{1}{u^{2}-1}+\frac{1}{u^{2}}+2\ln \left( 1-\frac{1}{u^{2}}\right) ,
\label{b21}
\end{equation}%
and, from Eq. (\ref{b18c}), we have defined a \textit{Hamaker temperature} as%
\begin{equation}
T_{\mathrm{H}}=\frac{A_{\mathrm{H}}}{36k_{B}}\ ,  \label{b22}
\end{equation}%
which depends on the material which constitutes the particles. In most
cases, $A_{\mathrm{H}}$ lies between $10^{-20\text{ }}$and $10^{-19\text{ }}%
\mathrm{J}$, i.e., $2\ k_{B}T\lesssim A_{\mathrm{H}}\lesssim 20\ k_{B}T$,\
where $T=298.15\ \mathrm{K}$. A typical value $A_{\mathrm{H}}=0.5\times
10^{-20\text{ }}\mathrm{J}$ ($=10\ k_{B}T$) yields $T_{\mathrm{H}}=100\ 
\mathrm{K}$, and thus - at room temperature - 
\begin{equation*}
\phi ^{\mathrm{H}}\left( \sigma +\delta \right) \approx \left\{ 
\begin{array}{c}
\begin{array}{ccc}
-2\ k_{B}T &  & \text{ \ \ if \ }\lambda =0.1\ \text{,}%
\end{array}
\\ 
\begin{array}{ccc}
-0.6\ k_{B}T\allowbreak &  & \text{if \ }\lambda =0.2\ \text{,}%
\end{array}
\\ 
\begin{array}{ccc}
-0.2\ k_{B}T\allowbreak &  & \text{if \ }\lambda =0.3\ \text{.}%
\end{array}%
\end{array}%
\right.
\end{equation*}

Using Eq. (\ref{b20}) for $Y_{\mathrm{\max }}$ together with the criteria in
Eq. (\ref{v5b}), one finds the approximate lower bound $T^{\mathrm{\min }}/%
\mathrm{K}$ for the applicability, respectively, of the linear, quadratic
and cubic approximations to the Mayer function, as reported in Table \ref%
{tab1:temp}.

Thus, whereas the linear approximation works only at high temperatures, the
quadratic one is already sufficient even at room temperature if $\lambda
\gtrsim 0.2$.

Unfortunately, analytical integration of the expression (\ref{b18}) is not
possible, and consequently no result for $t_{ij}^{\mathrm{eq}\left( \mathrm{H%
}\right) }$ can be obtained directly from $\phi _{ij}^{\mathrm{H}}\left(
r\right) $. Nevertheless, in order to get a rough approximation to $t_{ij}^{%
\mathrm{eq}\left( \mathrm{H}\right) }$, we propose an analytically
integrable interpolation of the correct behavior of $\phi _{ij}^{\mathrm{H}%
}\left( r\right) $ at short and large distances, i.e. 
\begin{equation}
\phi _{ij}^{\mathrm{H-interp}}\left( r\right) =-\frac{A_{ij}^{\mathrm{H}}}{36%
}\left\{ \frac{3}{2}\frac{\sigma _{i}{}\sigma _{j}{}}{\sigma _{ij}}\frac{1}{%
r-\sigma _{ij}}\exp \left( -\frac{r-\sigma _{ij}}{L}\right) +\frac{\ \sigma
_{i}^{3}\sigma _{j}^{3}}{r^{6}}\left[ 1-\exp \left( -\frac{r-\sigma _{ij}}{L}%
\right) \right] \right\} ,  \label{b21b}
\end{equation}%
where $\sigma _{ij}+\delta \leq r<+\infty $, and $L$ acts as a screening
length. When $\delta \simeq 0.1\sigma $, $L=0.108\sigma \approx \delta $
yields a satisfactory contact value, i.e., $\phi _{ij}^{\mathrm{H-interp}%
}\left( \sigma _{ij}+\delta \right) \approx \phi _{ij}^{\mathrm{H}}\left(
\sigma _{ij}+\delta \right) $. Using the linear approximation - valid at
high temperatures - one gets 
\begin{equation*}
t_{ij}^{\mathrm{eq}\left( \mathrm{H-interp}\right) }\approx \frac{1}{k_{B}T}%
\frac{A_{ij}^{\mathrm{H}}}{24}\frac{\sigma _{i}{}\sigma _{j}{}}{\sigma
_{ij}^{2}}\left[ E_{1}\left( \frac{\delta }{L}\right) +2e^{-\delta /L}\frac{L%
}{\sigma _{ij}}+\left( 1+\frac{\delta }{L}\right) e^{-\delta /L}\left( \frac{%
L}{\sigma _{ij}}\right) ^{2}\right] ,
\end{equation*}%
where $E_{1}(z)=\int_{z}^{+\infty }\frac{e^{-u}}{u}du$ is the exponential
integral. However, since the factors $L/\sigma _{ij}^{n}\approx \delta
/\sigma _{ij}^{n}$ refer to big particles, the leading term - at least
within the linear approximation - is 
\begin{equation}
t_{ij}^{\mathrm{eq}\left( \mathrm{H-interp}\right) }\approx \frac{3}{2}%
E_{1}\left( \frac{\delta }{L}\right) \left( \frac{T_{\mathrm{H}}}{T}\ \frac{%
\sigma _{i}{}\sigma _{j}{}}{\sigma _{ij}^{2}}\right) .  \label{b23}
\end{equation}

%\bigskip

\subsection{Polarizable hard spheres. Sutherland model}

Focusing only on the $r^{-6}$ part of the Hamaker potential, which
represents the long-distance polarizability, one could define a simpler
model, corresponding to a mixture of \textit{mesoscopic} HS with dispersion
attractions, called \textit{polarizable} hard spheres (PHS), i.e., 
\begin{equation}
\phi _{ij}^{\mathrm{PHS}}(r)=\left\{ 
\begin{array}{cc}
+\infty , & \qquad 0<r<\sigma _{ij} \\ 
-A_{ij}\ \sigma _{i}^{3}\sigma _{j}^{3}/r^{6} & r\geq \sigma _{ij},%
\end{array}%
\right.  \label{b14}
\end{equation}%
where the choice 
\begin{equation}
A_{ij}=\frac{A_{ij}^{\mathrm{H}}}{36}
\end{equation}%
ensures the mesoscopic size of the particles. If all particles are made up
of the same material substance, then $\phi _{ij}^{\mathrm{PHS}}(r)=-A\
\sigma _{i}^{3}\sigma _{j}^{3}/r^{6}$ for $r\geq \sigma _{ij}.$

The potential (\ref{b14}) may be regarded as a special case of the \textit{%
Sutherland model}, which represents rigid spheres which attract one another
according to an inverse-power law, i.e., $\phi ^{\mathrm{Sutherland}%
}(r)=-\epsilon \left( \sigma /r\right) ^{b}$ \ for $r\geq \sigma $ \ ($%
\epsilon >0$) \cite{Hirschfelder65}. Indeed, one could rewrite it as $\phi
_{ij}^{\mathrm{PHS}}(r)=-\epsilon _{ij}^{\mathrm{PHS}}\left( \sigma
_{ij}/r\right) ^{6}$, with $\epsilon _{ij}^{\mathrm{PHS}}=A\ \left( \sigma
_{i}\sigma _{j}/\sigma _{ij}^{2}\right) ^{3}$ for particles with the same
material composition.

The strength of this interaction - in the one-component case - is then given
by%
\begin{equation*}
Y_{\mathrm{\max }}\equiv -\beta \phi ^{\mathrm{PHS}}\left( \sigma \right) =%
\frac{T_{\mathrm{H}}}{T}\approx \frac{100\ \mathrm{K}}{T},
\end{equation*}%
after taking $T_{\mathrm{H}}\approx 100\ \mathrm{K}$. From the results
outlined in Eq.~(\ref{v5b}), a linearization of the Mayer function makes
sense for $T\gtrsim 1000\ \mathrm{K}$. A quadratic approximation is feasible
when $T\gtrsim 167\ \mathrm{K}$. Finally, the cubic approximation holds for $%
T\gtrsim 100\ \mathrm{K}$.

Therefore, in the multi-component case, one can safely adopt the cubic
approximation to $f_{ij}^{\mathrm{PHS}}(r)$ and perform the integration in
Eq. (\ref{b10}), obtaining 
\begin{equation}
t_{ij}^{\mathrm{eq}\left( \mathrm{PHS}\right) }=12\ \frac{T_{\mathrm{H}}}{T}%
\left( \frac{\sigma _{i}\sigma _{j}}{\sigma _{ij}^{2}}\right) ^{3}+15\left[ 
\frac{T_{\mathrm{H}}}{T}\left( \frac{\sigma _{i}\sigma _{j}}{\sigma _{ij}^{2}%
}\right) ^{3}\right] ^{2}+8\left[ \frac{T_{\mathrm{H}}}{T}\left( \frac{%
\sigma _{i}\sigma _{j}}{\sigma _{ij}^{2}}\right) ^{3}\right] ^{3}.
\end{equation}%
In the one-component case this expression reduces to%
\begin{equation*}
t^{\mathrm{eq}\left( \mathrm{PHS}\right) }=12\ \frac{T_{\mathrm{H}}}{T}%
+15\left( \frac{T_{\mathrm{H}}}{T}\right) ^{2}+8\left( \frac{T_{\mathrm{H}}}{%
T}\right) ^{3},
\end{equation*}%
For $T_{\mathrm{H}}\approx 100\ \mathrm{K}$, this results -- at $T\approx
300\ \mathrm{K}$ -- in a value $t^{\mathrm{eq}\left( \mathrm{PHS}\right)
}\simeq 5.96$, which corresponds in Baxter's parametrization to%
\begin{equation*}
\tau =\frac{1}{12t}\simeq 0.0\allowbreak 14\text{,}
\end{equation*}%
and lies well below the critical temperature of the SHS fluid, $\tau
_{c}=0.1133\pm 0.0005$ \cite{Miller0304}.

\section{$\allowbreak $Excluded volume depletion forces}

In general, the \textit{indirect, solvent-mediated,} solute-solute
interactions depend on both the solute-solvent and solvent-solvent ones, and
thus may be very difficult to evaluate \cite%
{Amokrane04,Louis02,Regnaut95,Amokrane98,Malherbe02}. We will now report
several very simplified cases.

\ Asakura and Oosawa (AO) \cite{Asakura5458}, and independently Vrij \cite%
{Vrij76}, first showed that two big (colloidal, or solute) particles,
immersed in a sea of small particles, feel a mutual attraction when their
surfaces get closer than the size of the smaller particles (\textit{%
depletion attraction}). This effect is an indirect attraction originating
from the interactions of the two big particles with the small ones of the
environment, even if these latter consist of, say, hard spheres. In mixtures
with neutral components, the small particles - hereafter referred to also as 
\textit{depletant} particles - may correspond, for example, to solvent
molecules, non-adsorbing polymer coils, or smaller colloidal particles.

Upon adding, for instance, polymers to a stable colloidal suspension, the
colloidal particles tend to aggregate. The polymer-induced depletion forces
between the colloidal particles can cause formation of colloidal crystals or
flocculation.

In the AO model, originally designed to describe colloid-polymer mixtures,
the big-big (colloid-colloid) interactions as well as the big-small ones are
modeled as excluded-volume HS interactions, while the small-small
interactions are assumed to be zero (ideal gas approximation, corresponding
to mutually interpenetrable, non-interacting depletant molecules). In
particular, polymer coils are assumed to have an effective HS diameter equal
to twice their radius of gyration.

Consider two big HS of species $i$ and $j$ at distance $r$, with radii $R_{i%
\text{ }}$and $R_{j}$, in a dilute suspension of depletant spheres of
species $0$, with radius $R_{0}$. The solute molecule $i$ produces a
spherical \textit{excluded-volume} region of radius $\sigma
_{i0}=R_{i}+R_{0} $ around itself where the centers of the depletant
particles cannot penetrate; this is also called the \textit{depletion zone}.
When the shortest distance $r-\left( R_{i}+R_{j}\right) =r-\sigma _{ij}$
between the surfaces of $i$ and $j$ becomes less than the diameter $\sigma
_{0}=2R_{0}$, the two depletion spheres surrounding $i$ and $j$ overlap and
the small particles are expelled from the region between the big molecules.
This implies that the thermal impact forces on the pair $i$ and $j$ from the
\textquotedblleft outside\textquotedblright\ are only partially compensated
by those from the \textquotedblleft inside\textquotedblright\ (see Fig. 8 of
Ref. \cite{Vrij76}). The depletion effect is due to this unbalanced pressure
difference, which pushes the big particles together, resulting in a net
attraction. From another point of view, the overlapping of excluded volumes
implies that the total \textit{free volume} accessible to small particles
increases, leading to a gain in the system entropy with a consequent
decrease of the Gibbs free energy. This trend to decrease free energy
produces an effective indirect attraction between the big spheres. AO and
Vrij \cite{Asakura5458,Vrij76} evaluated the \textit{HS-depletion }(HS-depl)%
\textit{\ }potential as%
\begin{equation}
\phi _{ij}^{\mathrm{HS-depl}}\left( r\right) =\left\{ 
\begin{array}{cc}
+\infty & \qquad 0<r<\sigma _{ij} \\ 
-\rho _{0}k_{B}T\ V_{ij}^{\mathrm{overlap}}\left( r\right) & \qquad \sigma
_{ij}\leq r\leq D_{ij} \\ 
0 & r>D_{ij},%
\end{array}%
\right.  \label{c7}
\end{equation}%
\begin{equation}
V_{ij}^{\mathrm{overlap}}(r)=\frac{\pi }{12}\frac{1}{r}\left(
D_{ij}-r\right) ^{2}\left[ 3D_{ii}D_{jj}-4D_{ij}\left( D_{ij}-r\right)
+\left( D_{ij}-r\right) ^{2}\right]  \label{c8}
\end{equation}%
where $\rho _{0}$ is the number density of the depletant molecules, $%
D_{ij}\equiv \sigma _{ij}+\sigma _{0}$, and $V_{ij}^{\mathrm{overlap}}(r)$
denotes the lens-shaped overlap volume of two spheres with radii $\sigma
_{i0}=\left( \sigma _{i}+\sigma _{0}\right) /2$ and $\sigma _{0j}=\left(
\sigma _{j}+\sigma _{0}\right) /2$, at distance $r$ (see Appendix). The
attraction increases linearly with temperature and with the concentration of
depletant particles. Since the AO model includes only HS interactions, the
corresponding depletion forces have a purely entropic origin. Finally, it
should be emphasized that the AO approximation is valid only for dilute
suspensions of depletant molecules, i.e., at low $\rho _{0}$ values;
formally, this last restriction can be removed by replacing $\rho_0$ by the
density of polymer in a large reservoir connected to the system.

The tail of $\phi _{ij}^{\mathrm{HS-depl}}\left( r\right) $ has a finite
range, equal to the diameter $\sigma _{0}$ of the depletant molecules. The
attraction strength can be estimated from 
\begin{equation}
Y_{ij,\mathrm{\max }}\equiv -\beta \phi _{ij}^{\mathrm{HS-depl}}\left(
\sigma _{ij}\right) =\eta _{0}\left( \allowbreak 1+\frac{3}{2}\frac{\sigma
_{i}\sigma _{j}}{\sigma _{ij}}\frac{1}{\sigma _{0}}\right) ,
\end{equation}%
where $\eta _{0}=\left( \pi /6\right) \rho _{0}\sigma _{0}^{3}$ is the
packing fraction of the depletant particles. Note that $Y_{ij,\mathrm{\max }%
} $ does not depend on temperature, but the attraction strength may be tuned
by varying $\eta _{0}$.

For one-component solutes, $Y_{ij,\mathrm{\max }}$ reduces to $Y_{\mathrm{%
\max }}=\eta _{0}\left( 1+1.5/\lambda \right) $ with $\lambda \equiv \sigma
_{0}/\sigma $. In this case - following again the criteria given in Eq. (\ref%
{v5b}) - the upper boundary $\eta _{0}^{\mathrm{\max }}$ for the
applicability of the linear, quadratic and cubic approximations to the Mayer
function given as a function of $\lambda $, respectively, is reported in
Table \ref{tab2:packing}.

The quadratic approximation result for the equivalent SHS model is 
\begin{eqnarray*}
t_{ij}^{\mathrm{eq}\left( \mathrm{HS-depl}\right) } &\approx &\frac{\eta _{0}%
}{2}\left[ \frac{\sigma _{i}\sigma _{j}}{\sigma _{ij}^{2}}\allowbreak
+\left( 1+\frac{1}{4}\frac{\sigma _{i}\sigma _{j}}{\sigma _{ij}^{2}}%
\allowbreak \right) \lambda _{ij}+\frac{1}{2}\lambda _{ij}^{2}+\frac{1}{12}%
\lambda _{ij}^{3}\right] \\
&&+\frac{1}{10}\left( \frac{\eta _{0}}{2}\right) ^{2}\left[ 9\left( \frac{%
\sigma _{i}\sigma _{j}}{\sigma _{ij}^{2}}\right) ^{2}\frac{1}{\lambda _{ij}}%
+16\frac{\sigma _{i}\sigma _{j}}{\sigma _{ij}^{2}}+\frac{4}{7}\left( \frac{13%
}{7}+4\frac{\sigma _{i}\sigma _{j}}{\sigma _{ij}^{2}}\right) \lambda _{ij}+%
\frac{17}{7}\lambda _{ij}^{2}+\frac{17}{63}\lambda _{ij}^{3}\right]
\end{eqnarray*}%
where $\lambda _{ij}\equiv \sigma _{0}/\sigma _{ij}$. As remarked, this
expression does not depend on $T$, since the solute-solvent interactions are
of purely HS character. If $\lambda _{ij}\ll 1$, then the leading terms are%
\begin{equation*}
t_{ij}^{\mathrm{eq}\left( \mathrm{HS-depl}\right) }\approx \left[ \frac{\eta
_{0}}{2}+\frac{8}{5}\left( \frac{\eta _{0}}{2}\right) ^{2}\right] \left( 
\frac{\sigma _{i}\sigma _{j}}{\sigma _{ij}^{2}}\right) \allowbreak +\frac{9}{%
10}\left( \frac{\eta _{0}}{2}\right) ^{2}\frac{1}{\lambda _{ij}}\left( \frac{%
\sigma _{i}\sigma _{j}}{\sigma _{ij}^{2}}\right) ^{2}.
\end{equation*}

Generalizing from the form of the quadratic approximation for general $%
\lambda _{ij}$, one expects the \emph{cubic} approximation to yield a result
of the form 
\begin{equation}
t_{ij}^{\mathrm{eq}\left( \mathrm{HS-depl}\right) }\approx C_{1}\left( \eta
_{0}\frac{\sigma _{i}\sigma _{j}}{\sigma _{ij}^{2}}\right) \allowbreak
+C_{2}\left( \eta _{0}\frac{\sigma _{i}\sigma _{j}}{\sigma _{ij}^{2}}\right)
^{2}+C_{3}\left( \eta _{0}\frac{\sigma _{i}\sigma _{j}}{\sigma _{ij}^{2}}%
\right) ^{3}.
\end{equation}

Several other studies of depletion forces, which go beyond the entropic HS
approach by taking into account more refined representions of the
solute-solvent and solvent-solvent interactions, are also available in the
recent literature \cite{Amokrane04,Louis02,Regnaut95,Amokrane98,Malherbe02}.

\bigskip

\section{Polymer-coated colloids or hairy spheres}

\ \ If the intermolecular attractive forces are strong enough, a colloidal
suspension phase-separates, or even flocculates or gels. As explained above,
stability against flocculation may be ensured by steric or charge
stabilization. In steric stabilization, the colloidal molecules are coated
with grafted polymers - the \textquotedblleft hair\textquotedblright\ \ -
which can prevent particles from coming sufficiently close to experience a
strong vdW-attraction.

However, changing the solvent or the temperature may turn the effective
interaction from repulsion (HS-behavior) to attraction \cite%
{Kruif89,Verduin95,Duits91}. When sterically-stabilized colloidal particles
are immersed in a \textit{good} solvent for the polymer brushes, the solutes
behave like HS, independently of temperature; this is the case, for example,
of silica particles covered with a layer of octadecyl chains, when immersed
in cycloexane. On the other hand, for each \textit{poor} solvent there
exists a Flory's \textit{theta-temperature}\ $T_{\theta }$ \cite{Flory53},
which is characteristic of the given solvent-polymer pair and has the
following property: the solute particles behave like HS at $T>T_{\theta }$,
whereas an attraction occurs at $T<T_{\theta }$. This occurs with e.g.\
silica particles with octadecyl chains, when dispersed in benzene. The term
\textquotedblleft $\theta $-solvent\textquotedblright\ indicates a poor
solvent at $T=T_{\theta }$.

These effects originate from a competition between polymer-solvent and
polymer-polymer interactions. First, the nature of the solvent influences
the polymer conformation. In fact, in a good solvent the interactions
between polymer elements - monomer units - and adjacent solvent molecules
are strongly attractive and thus predominate over possible intra-chain
polymer attractions. Consequently, the polymer will assume an
\textquotedblleft extended-chain\textquotedblright\ configuration, so as to
reduce the number of intra-chain contacts between monomer units.
Polymer-coated colloidal particles will have fully-extended hair and thus
the largest HS diameter possible, corresponding to the strongest
solute-solute repulsion.

In a poor solvent, on the other hand, the polymer-solvent attractions are
weak. Now it is the temperature that determines the solute-solute
interaction. At $T>T_{\theta }$ the hair will be fully-extended, as in good
solvents (HS behavior). At low temperatures $T<T_{\theta }$ the polymer
segments find their own environment more satisfying than that provided by
the solvent. This preference may produce more compact \textquotedblleft
globular\textquotedblright\ configurations, in which intra-chain
polymer-polymer contacts occur more frequently (\textquotedblleft curly
hair\textquotedblright ). In an alternative view, when two solute particles
are in close contact, a high number of polymer-polymer attractions is
favored by the interpenetration of the two polymeric layers.

In the literature on sterically-stabilized colloids \cite%
{Kruif89,Verduin95,Duits91}, the attractive part of the potential for
one-component \textit{hairy hard spheres} (HHS) - due to the polymer-polymer
interactions between surface layers of different particles - was described
by a SW, with a depth proportional to the (maximum) \textit{overlap volume }%
of the layers and temperature-dependent in analogy with the Flory-Krigbaum
model for polymer segments \cite{Flory50}. The SW width equals the length of
interpenetration of the stabilizing chains, whose maximum value coincides
with the thickness $\ell $ of polymeric layer.

For mixtures, we could consider the most direct extension of the
one-component SW model. In such a case, using Eq. (\ref{b10}), a SW
potential could immediately be mapped onto a SHS one:%
\begin{equation}
t_{ij}^{\mathrm{eq}\left( \mathrm{SW}\right) }=\left[ \exp \left( \beta
\epsilon _{ij}^{\mathrm{SW}}\right) -1\right] \ \frac{1}{3}\left[ \left(
1+\Delta _{ij}^{\mathrm{SW}}\right) ^{3}-1\right] .  \label{b12}
\end{equation}%
When the SW is very narrow $\left( \Delta _{ij}^{\mathrm{SW}}\ll 1\right) $,
one can approximate 
\begin{equation}
t_{ij}^{\mathrm{eq}\left( \mathrm{SW}\right) }\approx \left[ \exp \left(
\beta \epsilon _{ij}^{\mathrm{SW}}\right) -1\right] \ \Delta _{ij}^{\mathrm{%
SW}}.  \label{b12b}
\end{equation}

However, instead of a discontinuous SW model, we prefer to propose a
potential with a similar but continuous attractive tail of finite range,
i.e., 
\begin{equation}
\phi _{ij}^{\mathrm{HHS}}(r)=\left\{ 
\begin{array}{cc}
+\infty & \qquad r<\sigma _{ij} \\ 
-k_{B}T\ F(T)\ \rho _{\ell }\ V^{\mathrm{overlap}}(R_{i}+\ell ,R_{j}+\ell ,r)
& \qquad \sigma _{ij}\leq r\leq \sigma _{ij}+2\ell \\ 
0 & r>\sigma _{ij}+2\ell ,%
\end{array}%
\right.  \label{d2}
\end{equation}%
where 
\begin{equation}
F(T)=\left\{ 
\begin{array}{cc}
0 & \qquad T>T_{\theta } \\ 
T_{\theta }/T-1 & \qquad T<T_{\theta }%
\end{array}%
\right. .  \label{d3}
\end{equation}%
Here, we call $F(T)$ \textit{Flory's temperature-function}, and $\phi _{ij}^{%
\mathrm{HHS}}(r)$ is assumed to be proportional to the overlap volume
between polymeric layers of the two HHS at separation $r$, with $\rho _{\ell
}$ being a number density proportional to the polymer density in the
stabilizing layer.

Within the linear approximation, one finds for the equivalent SHS model 
\begin{equation*}
t_{ij}^{\mathrm{eq}\left( \mathrm{HHS}\right) }\approx F(T)\ \frac{\eta
_{\ell }}{2}\ \left[ \frac{\sigma _{i}\sigma _{j}}{\sigma _{ij}^{2}}%
\allowbreak +\left( 1+\frac{1}{4}\frac{\sigma _{i}\sigma _{j}}{\sigma
_{ij}^{2}}\allowbreak \right) \frac{2\ell }{\sigma _{ij}}+\frac{1}{2}\left( 
\frac{2\ell }{\sigma _{ij}}\right) ^{2}+\frac{1}{12}\left( \frac{2\ell }{%
\sigma _{ij}}\right) ^{3}\right] ,
\end{equation*}%
where $\eta _{\ell }\equiv \frac{\pi }{6}\rho _{\ell }\left( 2\ell \right)
^{3}$. Since the thickness $\ell $ is much smaller than the particle sizes,
one may expect - by analogy with the HS-depletion model - that the cubic
approximation reads%
\begin{equation}
t_{ij}^{\mathrm{eq}\left( \mathrm{HHS}\right) }\approx C_{1}\left[ F(T)\
\eta _{\ell }\frac{\sigma _{i}\sigma _{j}}{\sigma _{ij}^{2}}\right]
\allowbreak +C_{2}\left[ F(T)\ \eta _{\ell }\frac{\sigma _{i}\sigma _{j}}{%
\sigma _{ij}^{2}}\right] ^{2}+C_{3}\left[ F(T)\ \eta _{\ell }\frac{\sigma
_{i}\sigma _{j}}{\sigma _{ij}^{2}}\right] ^{3}.  \label{d5}
\end{equation}%
Note that, since $F(T)=0$ when $T>T_{\theta }$, then $\lim_{T\rightarrow
+\infty }F(T)=0$. Thus the form of $F(T)\ $ensures that the high-temperature
condition (\ref{b5}) is satisfied.

\section{Solvation forces. Gurney-Friedman model}

An indirect interaction between solute particles may also arise from
solvation. To picture solvation effects, Gurney \cite{Gurney53} and Frank
and Evans \cite{Frank45} introduced the physically intuitive concept of 
\textit{cosphere} or \textit{solvation layer}. One supposes that any
isolated solute particle is surrounded by some region in which the solvent
has different properties than the bulk solvent, since its structure is
markedly affected by the presence of the solute: some of the
solvent-solvent bonds have been broken by the introduction of the
\textquotedblleft foreign\textquotedblright\ particle. Clearly, such a
region has no well-defined boundary, but Gurney's model assumes that
significant effects come from only the few solvent molecules that are
directly next to the solute particles, i.e., in a spherical shell whose
thickness $\delta $ is taken to be a few solvent diameters or even the size
of only one solvent molecule (for water, a molecular diameter of $2.76$ \AA\ %
is acceptable). This picture was first applied to electrolyte solutions by
Friedman and coworkers \cite{Friedman717377}. In the ionic case, however,
the previous definition of cosphere, with the same thickness for every ionic
species, may be too restrictive, since the solvation region may extend even
outside the cosphere, as occurs for very small ions (Li$^{+}$, and
polyvalent ions such as Mg$^{2+}$, Ca$^{2+}$, etc.).

When two solute particles $i$ and $j$ approach sufficiently closely for
their solvation layers to overlap, some of the cosphere solvent is
displaced. Furthermore, the overlapping region contains solvent molecules
which are now affected by the combined force field of two solutes,\ so that
its structure might even differ from that of each isolated cosphere. The
whole process, in which the sum of the cosphere volumes is reduced by
overlap and the solvent relaxes to its normal bulk state, will be
accompanied by a Gibbs free energy change. If the solvent molecules in the
isolated solvation layers are in a state of lower free energy than those in
the bulk, the overlap of two cospheres with the consequent expulsion of
solvent gives rise to a free energy increase, and the resulting contribution
to the interaction between two solutes is repulsive. When the solvent
molecules in the solvation layers are in a state of higher free energy than
those in the bulk, the expulsion of solvent from the overlapping region
leads to a free energy decrease. In this case, both the free energy and the
disruption of solvent structure are minimized when two solute particles $i$
and $j$ are brought close together, causing a net $i-j$ attraction.

Because of the lack of knowledge about the properties of the solvent in the
solvation region, it is difficult to construct a detailed and physically
sound microscopic model of the effects described above. Adopting a heuristic
approach, Friedman and coworkers \cite{Friedman717377} proposed that the
free energy change accompanying the cosphere overlapping of two HS solute
particles $i$ and $j$ gives rise to the \textit{Gurney potential, }defined by%
\begin{equation}
\phi _{ij}^{\mathrm{Gurney}}(r)=A_{ij}(T,p)\ \frac{V^{\mathrm{overlap}%
}\left( R_{i}+\delta ,R_{j}+\delta ,r\right) }{v_{0}}.  \label{e1}
\end{equation}%
Here the Gurney parameter $A_{ij}$ is in general a function of temperature $%
T $ and pressure $p$ and represents the molar free energy of transfer of
solvent from the overlapping region of the $i-j$ cospheres to the bulk. As
previously discussed, $A_{ij}<0$ corresponds to attraction. Furthermore, $%
v_{0}$ is the mean volume of a solvent molecule, while the volume of solvent
returning to the bulk is given by the intersection volume of the cospheres
surrounding the two solute HS at distance $r$, namely $V^{\mathrm{overlap}%
}\left( R_{i}+\delta ,R_{j}+\delta ,r\right) $. The free-energy parameters $%
A_{ij}$ were determined numerically by fitting the model to experimental
data.

The close resemblance of $\phi _{ij}^{\mathrm{Gurney}}(r)$ to both $\phi
_{ij}^{\mathrm{HHS}}(r)$\ and $\phi _{ij}^{\mathrm{HS-depl}}\left( r\right) $
is apparent. By analogy one obtains immediately 
\begin{equation}
t_{ij}^{\mathrm{eq}\left( \mathrm{Gurney-solvation}\right) }\approx C_{1}%
\left[ E_{ij}(T,p)\ \frac{\sigma _{i}\sigma _{j}}{\sigma _{ij}^{2}}\right]
\allowbreak +C_{2}\left[ E_{ij}(T,p)\ \frac{\sigma _{i}\sigma _{j}}{\sigma
_{ij}^{2}}\right] ^{2}+C_{3}\left[ E_{ij}(T,p)\ \frac{\sigma _{i}\sigma _{j}%
}{\sigma _{ij}^{2}}\right] ^{3}.  \label{e2}
\end{equation}%
where%
\begin{equation}
\ E_{ij}(T,p)=\frac{1}{2}\frac{\left\vert A_{ij}(T,p)\right\vert }{k_{B}T}%
\left[ \frac{\frac{\pi }{6}\left( 2\delta \right) ^{3}}{v_{0}}\right]
\allowbreak .
\end{equation}%
This expression may be applied, in particular, to both cases of solvation
interactions - hydrophobic bonding and interactions between reverse micelles
- whose physical origin will be illustrated in the following.

\subsection{Hydrophobic interaction}

The \textit{hydrophobic interaction} (or \textit{hydrophobic bonding})
consists in the tendency shown by nonpolar portions of molecules or ions
with long nonpolar chains or aromatic groups - for example, surfactants,
phospholipids, glycerides, and dyestuffs - to aggregate in aqueous
solutions, partially or completely removing such parts from contact with the
solvent \cite{Lyklema00,Franks78,BenNaim80}. This \textit{hydrophobic
attraction} between nonpolar entities, which occurs exclusively in water,
has - to a large extent - an \textit{entropic} origin, related to the strong
tendency of water molecules to form hydrogen bonds and associate \cite%
{Kleman03}.

The physical mechanisms underlying the solvation forces are rather well
understood when the solvent is water. Polar molecules or polar groups of a
solute feel strong attraction towards water molecules, and thus are said to
be \textit{hydrophilic} (\textquotedblleft water-loving\textquotedblright ).
On the other hand, nonpolar molecules or nonpolar groups \textquotedblleft
dislike\textquotedblright\ water, and are called \textit{hydrophobic}
(\textquotedblleft water-hating\textquotedblright , or \textquotedblleft
water-avoiding\textquotedblright ).

The \textit{hydrophobic effect} means that nonpolar particles have an
extremely weak solubility in water, since inserting one of them - a noble
gas atom, a chlorine or oxygen molecule, a hydrocarbon molecule, etc. - into
water may actually lead to an increase of Gibbs free energy, i.e., $\Delta
G_{\mathrm{solution}}>0$. Indeed, the formation of a new cavity requires the
breaking of many water-water hydrogen bonds with a considerable $\Delta G>0$%
, which cannot be compensated by the small $\Delta G<0$ provided by the new
solute-water vdW interactions (the nonpolar solute cannot participate in the
formation of hydrogen bonds). Then, in order to get a further decrease of $G$%
, the water molecules close to the solute reorient themselves, so as to
create as many hydrogen bonds with adjacent water molecules as possible. The
result is the formation of a \textquotedblleft cage\textquotedblright\ - or
hydration layer - around the solute, with more rigid water-water bonds than
in the bulk. Such an additional \textit{ordering} in the solvent, brought
about by the introduction of a solute molecule, implies a significant
entropy decrease, $\Delta S<0$, and thus a strong positive contribution $%
-T\Delta S$ to the total $\Delta G=\Delta H-T\Delta S$ of solution at
constant $T$ and $p$. This explains why nonpolar particles are hydrophobic:
one gets $\Delta G_{\mathrm{solution}}>0,$ when the entropic contribution
dominates over the entalpic one $\Delta H$, which is usually small and may
be positive or negative. On the other hand, at higher temperatures the
solubility increases, and one may find $\Delta G_{\mathrm{solution}}<0.$ In
fact, $\Delta S$ becomes much smaller, because the molecular thermal motion
struggles more efficiently against the structure formation around a nonpolar
solute.

The same hydrophobic effect is responsible for the above-mentioned \textit{%
hydrophobic bonding}, where nonpolar parts of molecules or ions tend to
aggregate. In fact, the solvent molecules prefer mutual contacts over those
with the \textquotedblleft foreign\textquotedblright\ substance (solute),
while the aggregation of solutes reduces the total volume of their
\textquotedblleft cages\textquotedblright , minimizing the loss of entropy.
The hydrophobic interaction arises when overlap of hydration-layers occurs,
and becomes increasingly attractive as the distance between two solute
particles decreases.

This phenomenon in acqueous solutions of alcohols was studied by Friedman
and Krishnan \cite{Friedman73b}, who used a model potential containing a
repulsive term of the $r^{-9\text{ }}$form, plus an attractive Gurney term
given by Eq. (\ref{e1}), representing the overlap between cospheres. For the
sustances they considered, these authors found values of the Gurney
coefficient $A_{xx\text{ }}$in the range $-190$ to $ -60\text{ cal\ }\mathrm{%
mol}^{-1}$. The cosphere thickness $\delta $ was taken to be $2.76\ \mathrm{%
\mathring{A}}$, corresponding to one molecular layer of water.

Clark \textit{et al.} \cite{Clark77} investigated the same physical systems
with a more refined model, including a core potential based upon
Lennard-Jones potentials for individual atom-atom interactions and again a
Gurney term for the hydrophobic attraction. They found a minimum $\simeq
-0.5 $ $k_{B}T$ in their potentials of average force, that implies an
overall tendency for those alcohols to associate when in an aqueous
environment.

Hydrophobic bonding is very important in interface and colloid science. It
is often the driving force behind the way in which biomolecules organize
themselves and it is responsible for the formation of micelles, bimolecular
layers, and lamellar structures.

\subsection{Reverse micelles in water-in-oil microemulsions}

Molecules having both hydrophilic and hydrophobic parts are said to be 
\textit{amphiphilic} (\textquotedblleft dual-loving\textquotedblright , in
the sense of being both \textquotedblleft water-loving\textquotedblright\
and \textquotedblleft water-hating\textquotedblright ; from the Greek $%
\alpha \mu \varphi \iota $ = \textquotedblleft on both
sides\textquotedblright ). An important example is provided by relatively
short chain molecules with an ionizable or polar (thus hydrophilic)
head-group and a nonpolar (thus hydrophobic) tail, consisting of one or
several flexible hydrocarbon chain(s). Since these molecules can
significantly lower the surface tension of a solution, they are generically
called \textit{surfactants} or \textit{surface-active agents }\cite%
{Barrat03,Lyklema00,BenNaim80}. When immersed in water, the head-group may
become negatively or positively charged, or it remains polar with no net
electric charge. Accordingly, the surfactants are classified as \textit{%
anionic, cationic} or \textit{non-ionic}.

Clearly, it is the hydrophilic head-group that keeps a surfactant solute
dissolved in the water. The hydrophobic tail tends to avoid contact with
water and to seek, as far as possible, a nonaqueous environment. The longer
the hydrophobic tail, the poorer the solubility in water and, hence, the
greater the tendency of the surfactant to escape from the acqueous solution.
Consequently, as the solute concentration increases, a phase-separation may
occur. Alternatively, the surfactant molecules accumulate at interfaces
between water and other liquids or gases, or spontaneously \textit{%
self-assemble} into supramolecular aggregates of mesoscopic size that
minimize the number of contacts between water and hydrocarbon tails and
maximize the number of tail-tail interactions.

If a surfactant is added to pure water under atmospheric pressure, its
molecules first form a monolayer film at the water-air interface, with polar
heads pointing towards the water and tails exposed to the air. Above a
certain concentration of surfactant (\textit{critical micellization
concentration}), one finds an abrupt change in the properties of the
solution; in fact, now the solute particles in the bulk begin to form
supramolecular aggregates like micelles, planar lamellar bilayers, and
vesicles, whose size and shape are, to a large extent, determined by the
geometric properties of the surfactant molecules \cite{Barrat03,Lyklema00}.
These (\textit{direct})\textit{\ micelles} have a nearly spherical
structure, in which the head-groups are placed at the surface and are thus
exposed to the aqueous environment, whereas the nonpolar tails occupy the
interior of the micelle, avoiding any contact with water.

A surfactant, when added to a mixture of water and oil (an organic liquid
immiscible with water), forms monolayers at every water-oil interface.
Several disordered or partially ordered phases are possible, depending on
temperature and surfactant concentration. In particular, one can get a 
\textit{microemulsion}, which is a two-phase suspension of finely divided
droplets of \textit{oil in water }(O/W), or \textit{water in oil} (W/O),
depending on the relative concentration of the two liquids. Each droplet is
coated with a monolayer film of surfactant, which separates water from oil.
In W/O microemulsions one finds \textit{reverse} (or \textit{inverted}) 
\textit{micelles}, where the core is formed by a droplet of water, and the
surrounding surfactant molecules now have the head-groups inside the
micelle, in contact with water, while their hydrocarbon tails point towards
the outside oil phase \cite{Lemaire83,Roux85,Bouaskarne01,Amokrane97}.
Clearly, such flexible tails resemble the polymer-hair of sterically
stabilized colloids, but in reverse micelles the number of chains is lower
and thus a large amount of oil can penetrate the surfactant-layer. For the
sake of simplicity, we do not consider the possible presence of a
cosurfactant, which is generally an alcohol and mixes with surfactant in the
outside layer.

In most models for W/O microemulsions the suspending medium, containing
mainly oil, is treated as a continuous phase, and the reverse micelles are
represented as composed of two parts. The internal one, including both the
water droplet and surfactant head-groups, is assumed to be a spherical and
impenetrable core, with HS radius $R$. The external part consists of a
concentric, \textit{penetrable} spherical layer, with thickness equal to the
length $\ell $ of the aliphatic chains of surfactant in their fully-extended
conformation. Thus the total radius of a micelle is $R^{\prime }=R+\ell $.
Because of the flexibility of the chains, $\ell $ has sometimes been allowed
to depend on temperature, i.e., $\ell =\ell \left( T\right) $ \cite%
{Bouaskarne01}.

The short-ranged attraction between reverse micelles seems to be mainly
determined by the overlapping of the penetrable surfactant-layers during
collisions \cite{Lemaire83,Roux85,Amokrane97}. The interpenetration of the
aliphatic chains of the surfactant induces oil removal. Now, the partial
molar volume of oil inside the surfactant-layer is expected to be larger
than in the pure oil-phase (this effect is related to the volume of CH$_{2}$
and CH$_{3\text{ }}$groups in the aliphatic layers). A difference of $0.2$ 
\AA $^{3}$ seems to be sufficient to explain an interaction potential
compatible with light scattering experiments \cite{Roux85}.

Roux and Bellocq \cite{Roux85} proposed the simplest model for equal-sized
micelles, assuming that the interaction potential is proportional to the
overlap volume. The extension of their formula to mixtures is immediate.
According to our terminology, the resulting expression is equivalent to a
particular Gurney potential, i.e., 
\begin{equation}
\phi _{ij}^{\mathrm{rev.micel.}}(r)=\left\{ 
\begin{array}{cc}
+\infty & \qquad r<\sigma _{ij} \\ 
-k_{B}T\ \Delta \rho \ V^{\mathrm{overlap}}(R_{i}+\ell ,R_{j}+\ell ,r) & 
\qquad \sigma _{ij}\leq r\leq \sigma _{ij}+2\ell \\ 
0 & r>\sigma _{ij}+2\ell ,%
\end{array}%
\right.  \label{g1}
\end{equation}%
where $\Delta \rho $ is an adjustable parameter which depends only on the
oil.

More refined models also include a Hamaker-vdW attraction between water
cores. Electrostatic contributions and other more complicated terms have not
been considered in the present paper.

\section{Conclusions}

We have started from the problem of building up statistical mechanical
models for fluid mixtures of mesoscopic particles, like colloids, micelles
or globular proteins, by using very simple \textit{effective} potentials
containing a HS repulsion plus a short-ranged attractive tail that represent
the interaction between big particles after averaging out the microscopic
degrees of freedom related to much smaller molecules (solvent, added
polymers, etc.). The simplest tail corresponds to the highly idealized 
\textit{surface adhesion}, modeled through a $\delta _{+}$-term, of the SHS
potential. Since we are interested in multicomponent SHS fluids, and in the
past difficulties had been encountered in choosing their stickiness
parameters $t_{ij}$, the present paper has focused on
the relationship between $t_{ij}$ and particle sizes, i.e., on the possible
functional forms of $t_{ij}\left( T,\sigma _{i},\sigma _{j}\right) $. To
elucidate this issue we have regarded the SHS potentials as models that
may be derived, with some simplification, from more realistic interactions.
The idea of associating an \textit{equivalent} SHS representation to
a realistic interaction, by requiring the equality of the second virial
coefficients, is already known and widely accepted. We have chosen
this mapping based upon the virial equivalence because of its simplicity and
partial analytical tractability. Since $B_{2\text{ }}$yields only
the first correction of pressure with respect to ideality, one can
reasonably expect that the performance of the resulting SHS model should
worsen with increasing density. By requiring the equivalence of
quantities different from $B_{2\text{ }}$, one could obtain alternative
mappings and generally different values for the effective parameters,
leading to more successful results at high densities. An application of such
an idea was given for example in Ref. \cite{Amokrane97}, where the solvent molecules
were modeled as SHS by requiring the equivalence of structural properties,
i.e.\ the solvent-solvent structure factor and the coordination number.
Unfortunately, this approach does not admit an analytical treatment, and a
comparison with it goes beyond the scope of the present paper.

Our aim was to investigate the most interesting cases of short-ranged
attractive interactions for some paradigmatic physical systems. To provide a
physically sound basis for the choice of $t_{ij}$ and, in particular, its
dependence on the particle diameters, we have presented a detailed and
self-contained overview on several topics related to short-ranged attractive
interactions in real mixtures of neutral mesoscopic particles. We have
considered: i) the van der Waals or dispersion forces - direct interactions
- and three cases with indirect, solvent-mediated, attractions: ii)
depletion forces, iii) polymer-coated colloids, and iv) solvation forces (in
particular, hydrophobic bonding and interactions between reverse micelles in
water-in-oil microemulsions).

Due to obvious analytical difficulties, our analysis has been restricted to
the determination of the leading terms of $t_{ij}$. These have been
evaluated by series expansion of the Mayer function that appears in the
second virial coefficient. We have discussed in particular the linear,
quadratic and cubic approximations and their respective ranges of validity.

The main result is that - in almost all cases considered - the leading
contributions to $t_{ij}$ can be expressed as%
\begin{equation}
t_{ij}^{\mathrm{eq}\left( \mathrm{tail}\right) }\approx C_{1}\left[
M_{ij}(X,...)\ \frac{\sigma _{i}\sigma _{j}}{\sigma _{ij}^{2}}\right]
\allowbreak +C_{2}\left[ M_{ij}(X,...)\ \frac{\sigma _{i}\sigma _{j}}{\sigma
_{ij}^{2}}\right] ^{2}+\cdots ,  \label{co1}
\end{equation}%
where - in most scenarios - $X$ is the temperature $T$, and the property%
\begin{equation}
\lim_{T\rightarrow +\infty }M_{ij}(T,...)=0  \label{co2}
\end{equation}%
ensures that the \textit{high-temperature condition }of Eq.~(\ref{b5}) is
satisfied. In the case of the depletion attraction, $X$ coincides with the
packing fraction $\eta _{0}$ of the depletant particles. For hairy HS, both $%
T$ and $\eta _{\ell }$ are variables included in $M_{ij}$. As regards the
dependence of $t_{ij}$ on the big particle sizes - which was the basic
question of the present work - , it is remarkable that $t_{ij}$ can be
expressed as a simple polynomial in the variable $\left( \sigma _{i}\sigma
_{j}/\sigma _{ij}^{2}\right) $. A quadratic approximation may already be
sufficient. The only case in which the expression for $t_{ij}$ differs from
that given in Eq. (\ref{co1}) is the Sutherland model for polarizable HS,
which yields%
\begin{equation}
t_{ij}^{\mathrm{eq}\left( \mathrm{PHS}\right) }=12\ \frac{T_{\mathrm{H}}}{T}%
\left( \frac{\sigma _{i}\sigma _{j}}{\sigma _{ij}^{2}}\right) ^{3}+15\left[ 
\frac{T_{\mathrm{H}}}{T}\left( \frac{\sigma _{i}\sigma _{j}}{\sigma _{ij}^{2}%
}\right) ^{3}\right] ^{2}+8\left[ \frac{T_{\mathrm{H}}}{T}\left( \frac{%
\sigma _{i}\sigma _{j}}{\sigma _{ij}^{2}}\right) ^{3}\right] ^{3}.
\end{equation}%
It is pleasant that even here we find powers of the same \textit{basic
size-dependent factor}, $\left( \sigma _{i}\sigma _{j}/\sigma
_{ij}^{2}\right) $. Note that this factor has the property that the sticky
attraction vanishes when at least one of the two particles $i$ and $j$
becomes a point, i.e., it satisfies the \textit{point-limit} \textit{%
condition} 
\begin{equation}
\lim_{\sigma _{i}\text{ or }\sigma _{j}\rightarrow 0\text{ }}t_{ij}(T,\sigma
_{i},\sigma _{j})=0.  \label{r2}
\end{equation}%
This condition would be expected to hold for any interaction of \
\textquotedblleft adhesive\textquotedblright\ type which in the limit
involves a particle surface of vanishing area \cite{Barboy75}.

\begin{figure}[hbtp]
\begin{center}
\includegraphics[angle=0,width=6cm]{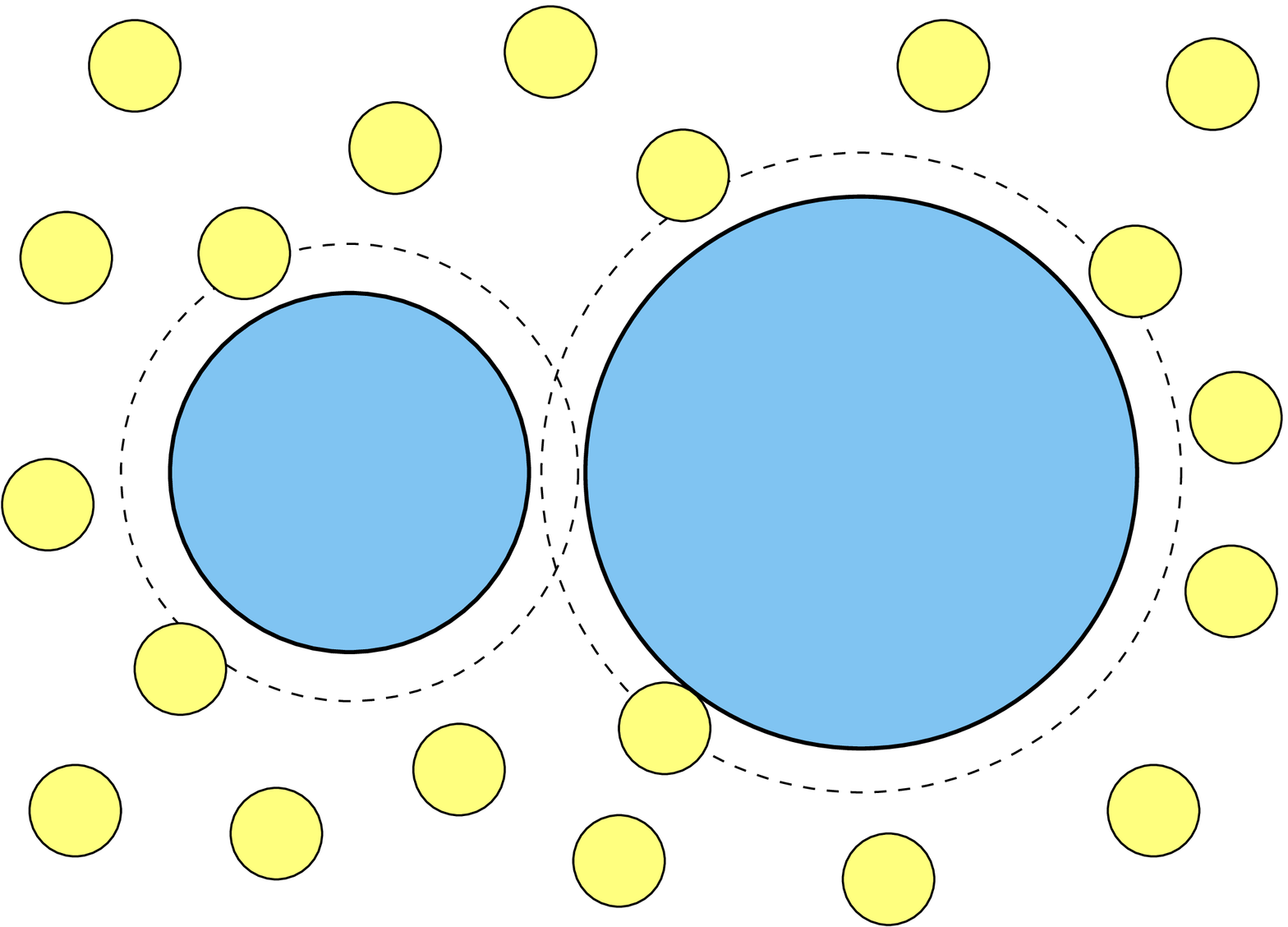} \hskip1.5cm %
\includegraphics[angle=0,width=6cm]{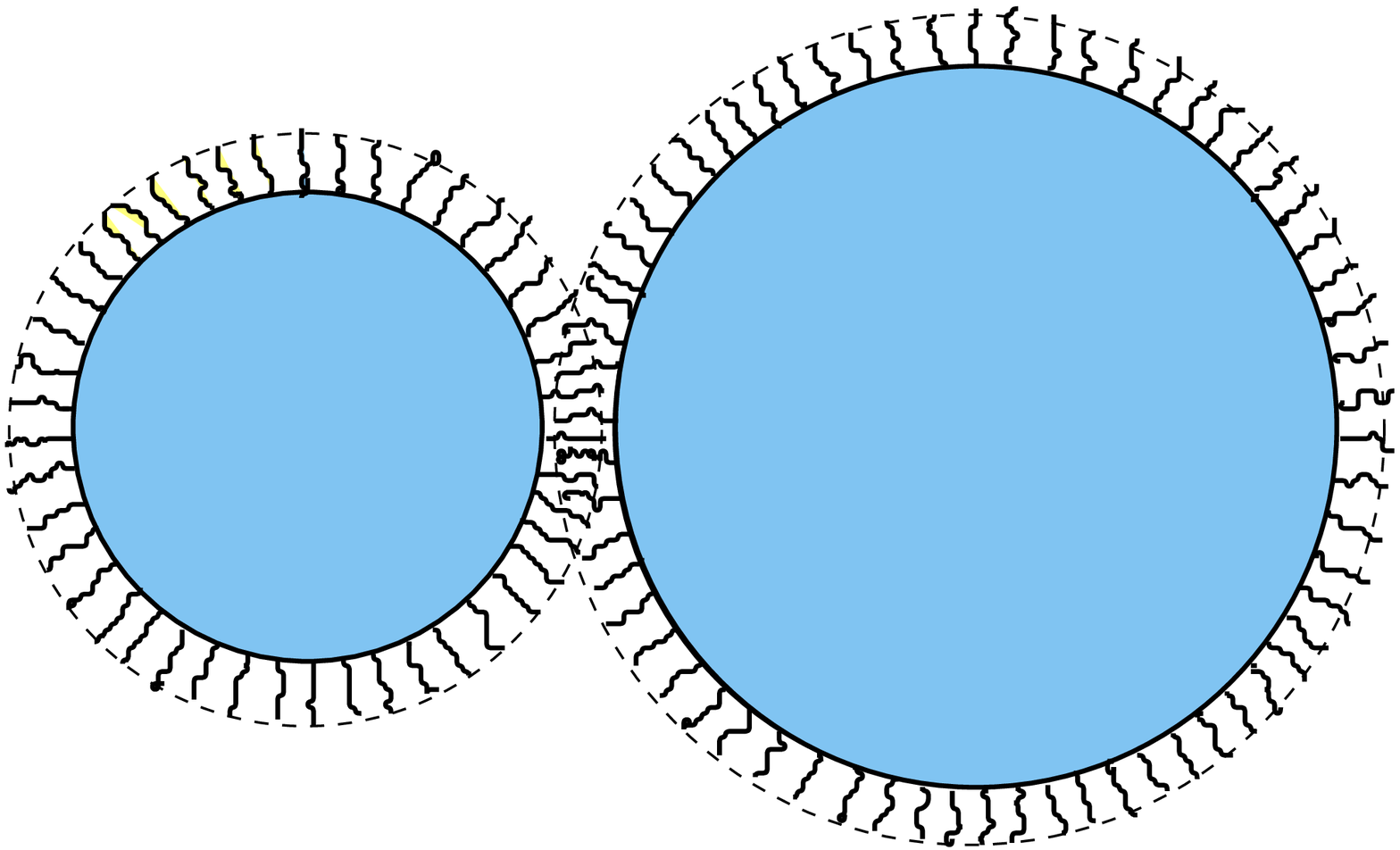}\\[0pt]
\vskip1.5cm \includegraphics[angle=0,width=6cm]{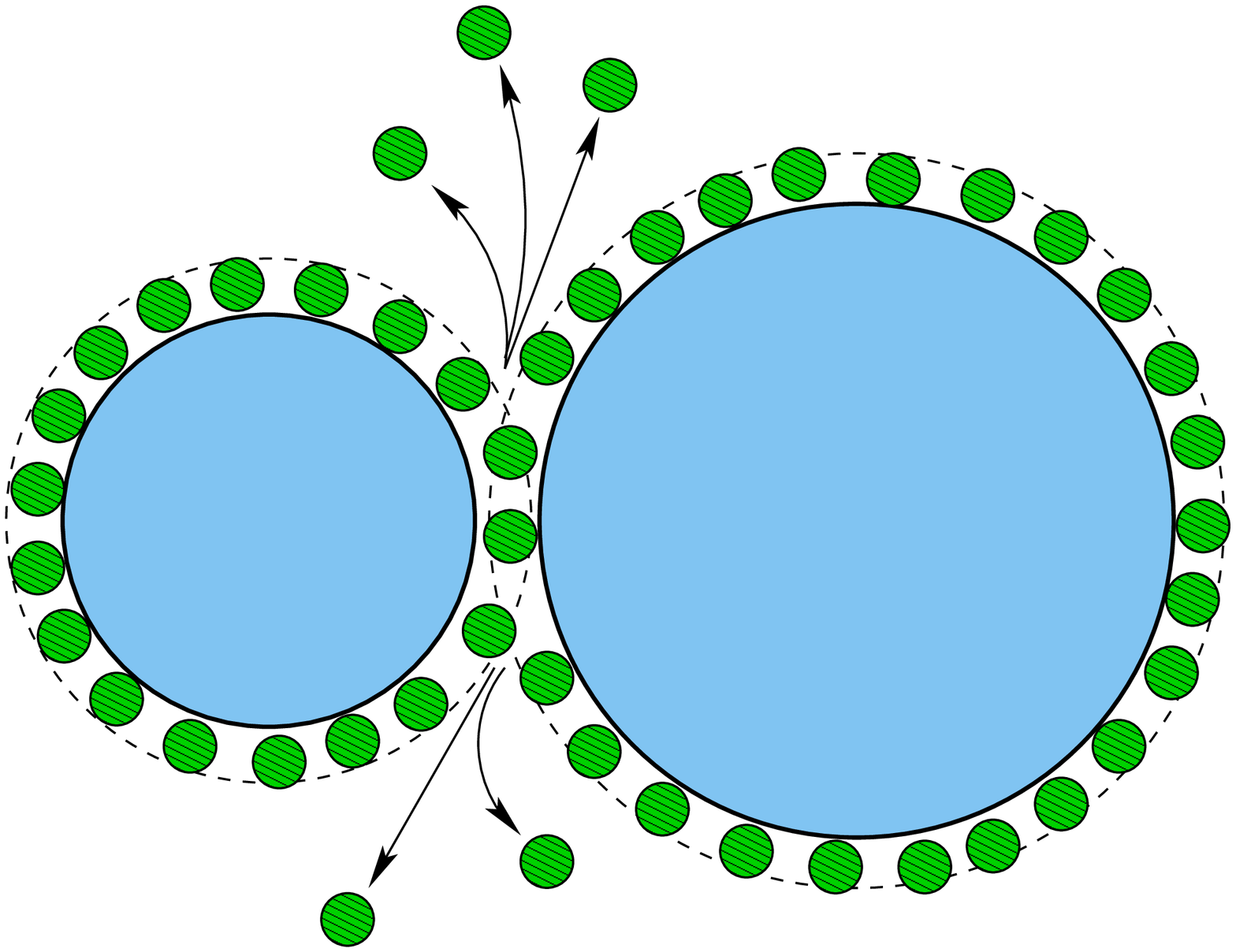} \hskip1.5cm %
\includegraphics[angle=0,width=6cm]{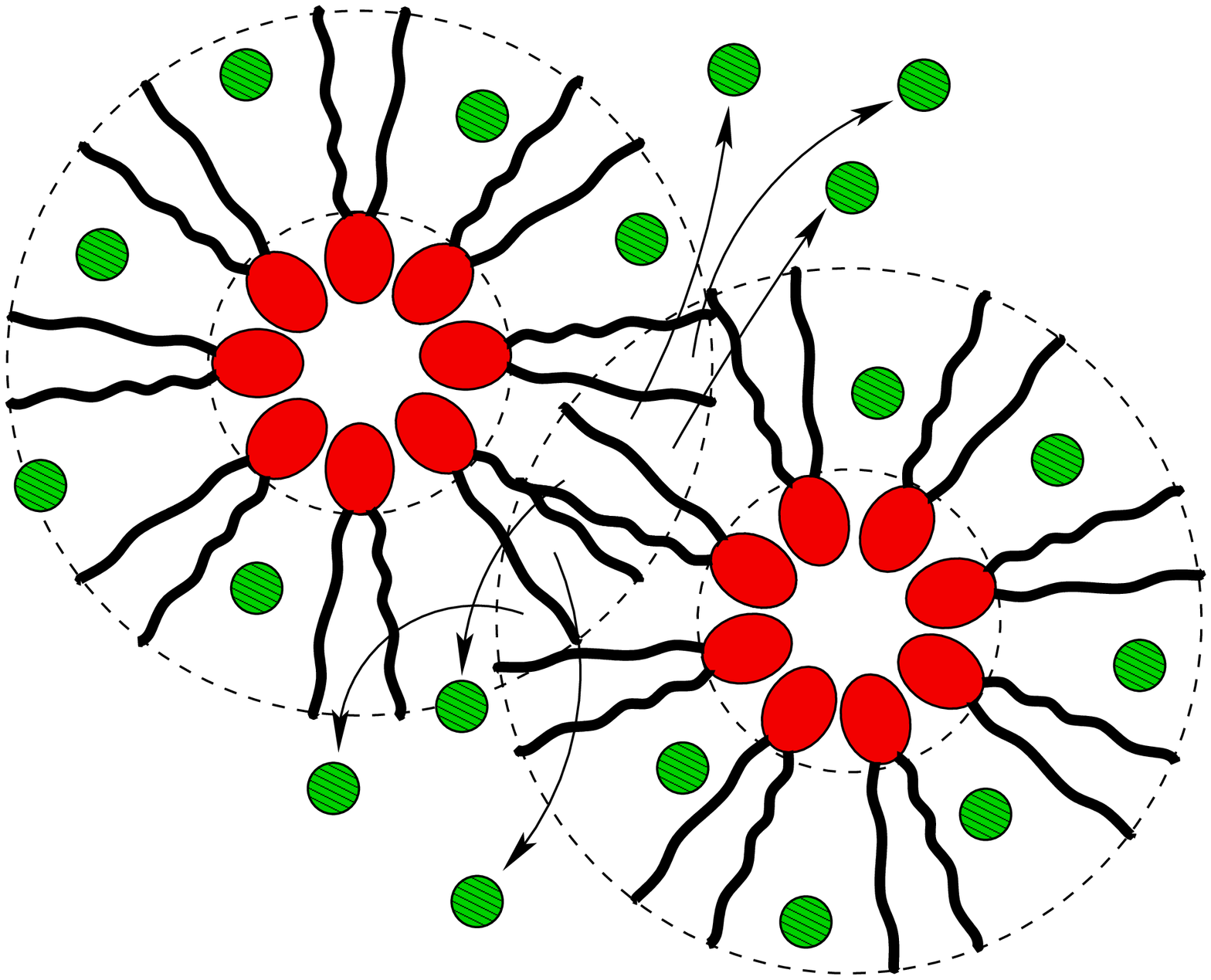}
\end{center}
\caption{ Schematic representation of some systems described by SHS
equivalent models in the text. \textit{Upper left panel}: Excluded-volume
depletion attraction between big spheres (solutes) in a sea of smaller
depletant particles (light-grey). The dashed curves represent the
excluded-volume regions where the centers of the depletant particles cannot
penetrate. \textit{Upper right panel}: Polymer-coated colloids or hairy
spheres, when their sterically-stabilizing layers of grafted polymers
overlap. \textit{Lower left panel}: Overlap of solvation layers
('cospheres') in the Gurney-Friedman model, and expulsion of solvent from
the overlapping region. The small spheres (dark-grey) represent solvent
molecules. \textit{Lower right panel}: Interaction between reverse micelles
in water-in-oil microemulsions. In each micelle the internal dashed curve
indicates the impenetrable core, formed by a droplet of water where the
head-groups of the surfactant molecules are immersed. The region between the
core and the external dashed curve is the penetrable part of the micelle,
corresponding to the hydrocarbon tails and containing some oil molecules
(small spheres). The micellar attraction is mainly due to oil removing from
the overlapping region and its transfer to the bulk. }
\label{fig1}
\end{figure}

The similarity among most of the resulting expressions for $t_{ij}^{\mathrm{%
eq}\left( \mathrm{tail}\right) }$ can easily be understood. In fact, most of
the different short-ranged attractions considered here have been explained -
by different authors - in terms of quite similar models, where mesoscopic
particles are represented as hard spheres with a surrounding, concentric
spherical \textit{layer} (see Figure \ref{fig1}). In the AO case this layer
is characterized somewhat indirectly by the fact that the density of the
centers of mass of the depletant polymers is zero inside the layer but has a
nonzero value outside it. In the other cases, the layer has a clearer
physical reality. For polymer-coated colloids, for example, the layer is the
polymeric film grafted on the colloidal surface. In the hydrophobic bonding
the layer is formed by the solvation water molecules. For reverse micelles,
the core comprising the water droplet and the polar heads of surfactants is
surrounded by a layer made up of hydrocarbon tails of surfactants plus a
certain quantity of oil.

It is very appealing, and somewhat surprising, that the factor $\sigma
_{i}\sigma _{j}/\sigma _{ij}^{2}$ appears even in our result for the Hamaker
potential, which refers to a direct interaction where no spherical layer
around the solutes is involved. Note that this dependence on the particle
diameters is clearly due to the Hamaker integration, since for point-dipoles
at the centre of hard spheres (polarizable HS) we have found a different
factor, i.e., $\left( \sigma _{i}\sigma _{j}/\sigma _{ij}^{2}\right) ^{3}.$

We remark that the models we have considered can be divided into two different
classes. The first one includes the two models of dispersion forces (Hamaker
potential and polarizable HS), with the common feature of having an
attractive $r^{-6}$ tail. The second class refers to the solvent-mediated
attractions (depletion effects, polymer-coated colloids, solvation forces).
Here, we have reported the simplest examples, which can be regarded as
variants of one single model: hard spheres with a penetrable concentric
spherical layer (`cosphere', in a wider sense). As a consequence, since the
attraction depends on the volume overlap of the cospheres, the potentials of
all these models are `truncated', i.e.\ they are rigorously zero beyond some
characteristic distance.

The main difference between the above mentioned classes -- infinite
tail in the first, finite tail in the second -- might suggest
that the idea of representing realistic potentials by an equivalent SHS
model is justifiable for the second class, but somewhat more questionable
when the tail is infinite. In particular, since a proper treatment of long
tails is essential for thermodynamics, the SHS-mapping of the Hamaker and
Sutherland potentials might introduce some qualitative differences in such a
kind of properties. This viewpoint is certainly correct and 
Hamaker and Sutherland potentials should be appropriately 
distinguished from the remaining models of this paper. In fact, an `exact'
treatment of all these models  would surely yield very different thermodynamic and structural
predictions. Nevertheless, in our context the $B_{2}$-mapping onto SHS 
can be expected to yield a representation of realistic interactions
that is simple, analytically tractable, and reliable in appropriate
regimes, at low and intermediate densities.

In a companion paper \cite{Sollich06}, we have applied a perturbative
approach to the solution of the polydisperse SHS model within the
Percus-Yevick approximation. The suggestions put forward in the present
paper regarding the relationship between stickiness and size, could
help to improve the necessary input to that kind of scheme.

In conclusion, the present paper suggests - for multicomponent SHS models -
the expression for $t_{ij}$ given by Eq. (\ref{co1}) as a simple choice that
is physically justified by its relation to the above-mentioned models of
real interactions. Clearly, Eq. (\ref{co1}) is an \textit{approximate}
result, but we believe that it correctly includes the \textit{leading terms}
of the dependence of $t_{ij}$ on the particle sizes. 
In spite of the rather drastic approximations used here, this could
be useful with the rationale of having a simple and tractable representation
of rather complex interactions, at the simplest possible level of description.

\acknowledgments
We are grateful to Andr\'es Santos and Sa\"id Amokrane for enlightening
discussions and critical reading of the manuscript. The work in Italy was
supported by a MIUR-COFIN 2006-2007 grant (\textit{Colloidal mixtures,
globular proteins and liquid crystal-like phases of biopolymers}). 
\appendix

\section{Volume overlap between spheres}

The volume of the intersection between HS with radii $a$ and $b$, at
distance $r$, is 
\begin{equation}
V^{\mathrm{overlap}}(a,b,r)=\left\{ 
\begin{array}{cc}
\left( 4\pi /3\right) \min \left( a^{3},b^{3}\right) & \qquad 0<r<\left\vert
a-b\right\vert , \\ 
\begin{array}{c}
\\ 
\frac{\pi }{12}\left[ -3\left( a^{2}-b^{2}\right) ^{2}\frac{1}{r}+8\left(
a^{3}+b^{3}\right) -6\left( a^{2}+b^{2}\right) r+r^{3}\right] \\ 
\end{array}
& \qquad \left\vert a-b\right\vert <r<a+b, \\ 
0 & r>a+b.%
\end{array}%
\right.  \label{a1}
\end{equation}
For $\left\vert a-b\right\vert <r<a+b$ this expression can conveniently be
rewritten as 
\begin{equation}
V^{\mathrm{overlap}}(a,b,r)=\frac{\pi }{12}\left[ 12ab\left( r-a-b\right)
^{2}+4\left( a+b\right) \left( r-a-b\right) ^{3}+\left( r-a-b\right) ^{4}%
\right] \ \frac{1}{r}.  \label{a2}
\end{equation}%
Taking $a=\sigma _{i0}=\left( \sigma _{i}+\sigma _{0}\right) /2$ and $%
b=\sigma _{0j}=\left( \sigma _{j}+\sigma _{0}\right) /2$, one gets $V^{%
\mathrm{overlap}}(\sigma _{i0},\sigma _{j0},r)=V_{ij}^{\mathrm{overlap}}(r)$
of Eq. (\ref{c8}).

\bigskip 
%%%%%%%%%%%%%%%%%%%%%% Bibliography %%%%%%%%%%%%%%%%%%%%%%%%%%%%%%%%%%%%
%\references

\newpage

%%%%%% Table 1 %%%%%%%%%%%%%%%%%%%
\begin{table}[tbp]
\begin{tabular}{lcccccc}
\hline
\text{Approximation} & $~~~~~~~$ & $\lambda=0.1$ & $~~~~~~~$ & $\lambda=0.2$
& $~~~~~~~$ & $\lambda=0.3$ \\ \hline
\text{linear} &  & 6300 &  & 1800 &  & 700 \\ 
\text{quadratic} &  & 1050 &  & 300 &  & 117 \\ 
\text{cubic} &  & 630 &  & 180 &  & 70 \\ \hline
\end{tabular}%
\caption{Approximate lower bound $T^{\mathrm{min}}/K$ for the applicability
of the linear, quadratic, cubic approximation to the Mayer function $f_{ij}$%
, as a function of the parameter $\protect\lambda=\protect\sigma_0/\protect%
\sigma$ (see text). }
\label{tab1:temp}
\end{table}

\vskip3.0cm

%%%%%% Table 2 %%%%%%%%%%%%%%%%%%%
\begin{table}[tbp]
\begin{tabular}{lcccccc}
\hline
\text{Approximation} & $\lambda=0.1$ & $\lambda=0.3$ & $\lambda=0.5$ & $%
\lambda=0.7$ & $\lambda=0.9$ & $\lambda=1.0$ \\ \hline
\text{linear} & 0.006 & 0.017 & 0.025 & 0.03 & 0.0375 & 0.04 \\ 
\text{quadratic} & 0.04 & 0.10 & 0.15 & 0.19 & 0.23 & 0.24 \\ 
\text{cubic} & 0.06 & 0.17 & 0.25 & 0.32 & 0.38 & 0.4 \\ \hline
\end{tabular}%
\caption{Approximate upper bound $\protect\eta_0^{\mathrm{max}}$ for the
applicability of the linear, quadratic, cubic approximation to the Mayer
function $f_{ij}$, as a function of $\protect\lambda$ defined above. }
\label{tab2:packing}
\end{table}

\newpage

%%%%%%%%%%%%%%%%%%%%%%%%%%%%%%%%%%% FIGURE %%%%%%%%%%%%%%%%%%%%%%%%%%%%%%


\begin{thebibliography}{99}
\bibitem{note1} $\ $ Taking the limit $w\rightarrow 0$ of the function $%
\varphi _{w}(x)=w^{-1}$ for $x_{0}<x<x_{0}+w$, and $=0$ elsewhere, one gets
the asymmetric Dirac delta distribution $\delta _{+}(x)$, which has the
property: $\int_{A}^{B}dx\ F(x)\delta _{+}(x-x_{0})=F\left( x_{0}^{+}\right)
,$ if $A\leq x_{0}<B$, and \ $=0,$ if $x_{0}<A$ or $x_{0}\geq B$.
\end{thebibliography}
\end{document}